\documentclass[twoside]{article}
\usepackage{fleqn,espcrc2}
\usepackage{epsfig}

\usepackage{graphicx}

\title{Matter Effects in Upward-Going Muons and Sterile Neutrino Oscillations}
\author{
\begin{center}
{\bf The MACRO Collaboration  } \\
\nobreak\bigskip\nobreak
M.~Ambrosio$^{12}$,
R.~Antolini$^{7}$,
G.~Auriemma$^{14,a}$,
D.~Bakari$^{2,17}$,
A.~Baldini$^{13}$,
G.~C.~Barbarino$^{12}$,
B.~C.~Barish$^{4}$,
G.~Battistoni$^{6,b}$,
Y.~Becherini$^{2}$,
R.~Bellotti$^{1}$
% ...
\\
\footnotesize
1. Dipartimento di Fisica dell'Universit\`a di Bari and INFN, 70126
Bari,  Italy \\
2. Dipartimento di Fisica dell'Universit\`a di Bologna and INFN,
 40126 Bologna, Italy \\
3. Physics Department, Boston University, Boston, MA 02215,
USA \\
4. California Institute of Technology, Pasadena, CA 91125,
USA \\
5. Department of Physics, Drexel University, Philadelphia,
PA 19104, USA \\
6. Laboratori Nazionali di Frascati dell'INFN, 00044 Frascati (Roma),
Italy \\
7. Laboratori Nazionali del Gran Sasso dell'INFN, 67010 Assergi
(L'Aquila),  Italy \\
8. Depts. of Physics and of Astronomy, Indiana University,
Bloomington, IN 47405, USA \\
9. Dipartimento di Fisica dell'Universit\`a dell'Aquila  and INFN,
 67100 L'Aquila,  Italy \\
10. Dipartimento di Fisica dell'Universit\`a di Lecce and INFN,
 73100 Lecce,  Italy \\
11. Department of Physics, University of Michigan, Ann Arbor,
MI 48109, USA \\
12. Dipartimento di Fisica dell'Universit\`a di Napoli and INFN,
 80125 Napoli,  Italy \\
13. Dipartimento di Fisica dell'Universit\`a di Pisa and INFN,
56010 Pisa,  Italy \\
14. Dipartimento di Fisica dell'Universit\`a di Roma ``La Sapienza and INFN,
 00185 Roma,   Italy \\
15. Physics Department, Texas A\&M University, College Station,
TX 77843, USA \\
16. Dipartimento di Fisica Sperimentale dell'Universit\`a di Torino and INFN,
 10125 Torino,  Italy \\
17.  L.P.T.P., Faculty of Sciences, University Mohamed I, B.P. 524 Oujda,
Morocco \\
18. Dipartimento di Fisica dell'Universit\`a di Roma Tre and INFN, 00146
Roma,   Italy \\
\end{center}
}
\begin{document}
\begin{abstract}
The angular distribution of upward-going muons
produced by atmospheric neutrinos in the rock
below the MACRO detector show anomalies in good
agreement with two flavor
$\nu_{\mu}\rightarrow\nu_{\tau}$ oscillations with maximum mixing and $\Delta m^2$ around
$0.0024$ $eV^2$.
Exploiting the dependence of magnitude of the matter effect on
oscillation channel, and using a set of 809 upward-going muons observed
in MACRO, we show  that the two flavor $\nu_{\mu}\rightarrow\nu_{s}$ oscillation
is disfavored with 99\% C.L. with respect to  $\nu_{\mu}\rightarrow\nu_{\tau}$.
\end{abstract}
\maketitle

\section{INTRODUCTION}
\footnotetext{
$a$ Also Universit\`a della Basilicata, 85100 Potenza,  Italy - \\
$b$ Also INFN Milano, 20133 Milano, Italy\\
$c$ Also Istituto TESRE/CNR, 40129 Bologna, Italy\\
$d$ Also Scuola Normale Superiore di Pisa, 56010 Pisa, Italy\\
$e$ Also Universit\`a di Trieste and INFN, 34100 Trieste, Italy\\
$f$ Also Dip. di Energetica, Universit\`a di Roma,  00185 Roma,  Italy \\
$g$ Also Institute for Nuclear Research, Russian Academy
of Science, 117312 Moscow, Russia \\
$h$ Also Institute for Space Sciences, 76900 Bucharest, Romania \\
$i$ The Colorado College, Colorado Springs, CO 80903, USA\\
$l$ Also Dept. of Physics, Pennsylvania State University,
    University Park, PA 16801, USA\\
$m$ Also Dipartimento di Fisica dell'Universit\`a della Calabria, Rende (Cs), It
aly \\
$o$ Also Dipartimento di Scienze e Tecnologie Avanzate,
Universit\`a  del Piemonte Orientale, Alessandria, Italy \\
$p$ Also U. Minn. Duluth Physics Dept., Duluth, MN 55812 \\
$q$ Also Dept. of Physics, MIT, Cambridge, MA 02139 \\
$r$ Also Intervideo Inc., Torrance CA 90505 USA \\
$s$ Also Resonance Photonics, Markham, Ontario, Canada\\
$*$ Corresponding author E-mail: ronga@lnf.infn.it}

Neutrino oscillations\cite{Pontecorvo1957} were first suggested by B. Pontecorvo
in 1957 after the discovery of the $K^{0} \leftrightarrow \overline{K^{0}}$ transitions.
Subsequently, evidence for the existance of neutrino
oscillation in nature has been provided by the Superkamiokande, Soudan2 and
MACRO experiments, each of which has presented data which strongly
favor atmospheric neutrino oscillations, in the form of $\nu_{\mu}$ disappearance \cite{Kajita:1998bw}.

 The two neutrino oscillation probability in vacuum    is given by:
\begin{equation}
   P(\nu_\ell \rightarrow \nu_{\ell^\prime\neq\ell}) =
 \sin^2 2\theta \,
\sin^2 \left[ 1.27 \: \Delta m^2
\frac{L}{E} \right]
\label{equ5}
\end{equation}
where
$\Delta m^2=m_{1}^2-m_{2}^2  (\mbox{ eV}^2) , L(\mbox{km}), E(\mbox{GeV})$,
 $\theta$ is the mixing angle and L is the path length between the neutrino
production point and the location at which the neutrino flavor is measured.
This simple relation should be modified when a  neutrino propagates through matter and when there
is a difference in the interactions of the two neutrino flavors with matter \cite{Matter}.
The neutrino weak potential  in matter is:
\begin{equation}
V_{\rm weak}=\pm\,\frac{G_F n_B}{\sqrt2}\times
\cases{-Y_n+2Y_e&for $\nu_e$,\cr
-Y_n&for $\nu_{\mu,\tau}$,\cr
0&for $\nu_s$,\cr}
\label{eq4}
\end{equation}
where the upper sign refers to neutrinos, the lower sign to
antineutrinos, $G_F$ is the Fermi constant, $n_B$ the baryon density,
$Y_n$ the neutron and $Y_e$ the electron number per baryon (both about
1/2 in common matter).
The weak potential in matter produces a phase shift that will  modify the neutrino
oscillation probability if the oscillating neutrinos have different interactions with matter.
Therefore, the matter effect could help to discriminate between different neutrino oscillation channels.
According  to equation, (~\ref{eq4})  matter effects in the Earth could be
important for
$\nu_{\mu}\rightarrow\nu_{e}$  and for the $\nu_{\mu}\rightarrow\nu_{s}$ oscillations, while
for  $\nu_{\mu}\rightarrow\nu_{\tau}$ oscillations there is no matter effect.
For particular values of the oscillation parameters the matter effect increases
the oscillation probability, leading to 'resonances' (e.g., the MSW effect).

$\nu_{\mu}\rightarrow\nu_{s}$ oscillations have been suggested \cite{sterile}
to explain some features of the atmospheric neutrino anomaly.
Under most current models, a fourth (sterile) neutrino is necessary to explain all the 
reported neutrino anomalies (solar,
atmospheric and LSND \cite{Mills:2001tq} ). Matter effects are important
\cite{sterile} when $E_\nu/|\Delta m^2| \ge 10^3$  GeV/eV$^2$,
  therefore in particular for high energy events. The
primary  purpose of this letter is to compare the MACRO high energy neutrino events
sample with the predictions, considering matter effects in the case of
 $\nu_{\mu}\rightarrow\nu_{s}$ oscillations. 
In MACRO, neutrino oscillation is observed in three different
event topologies, having different characteristic ranges of parent
neutrino energies. So-called $Up$ $Through$ events \cite{Ahlen:1995av} are
associated with muons which penetrate the entire detector.  The parent
neutrinos in these events have a median neutrino energy around
50 GeV. $Internal$ $Up$ events and $Internal$ $Down$ events, together with
$Up$ $Stop$ events, \cite{Macro00} are associated with muons having a track
terminus located within the MACRO detector. The parent neutrinos in
these events have a significantly lower median energy, of around 4 GeV.
In this paper, we focus on the high energy ($Up$ $Through$) data sample.
A similar analysis has been
recently published by the Superkamiokande collaboration \cite{Fukuda:2000np}.

\section{DATA ANALYSIS}
The MACRO detector \cite{Ahlen:1993pe}.
is located in the Hall B of the Gran Sasso
Laboratory, with a minimum rock overburden of 3150 hg/cm$^2$.
It is in the general form of a large rectangular box,
76.6~m~$\times$~12~m~$\times$~9.3~m, divided
longitudinally into six supermodules, and vertically into a lower
part (4.8 m high) and an upper part (4.5 m high). The
active detection elements are planes of streamer tubes for tracking
and of liquid scintillators for fast timing. The lower half of
the detector is filled
with trays of crushed rock absorbers alternating with streamer tube
planes, while the upper part is open and contains
electronics racks and work areas.
\begin{figure} [th]
\begin{center}
\mbox{
\epsfig{file=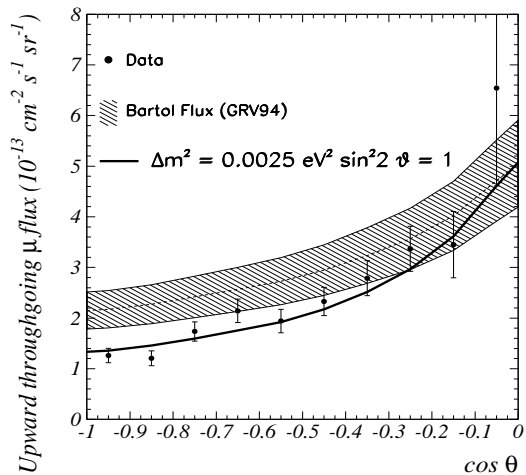,width=7.5cm,height=7cm}}
\caption {\label{angleflux}\small Zenith distribution of the flux of upgoing
muons with energy greater than 1 GeV for the combined MACRO
data. The shaded region shows the expectation for no oscillations
with the 17\% normalization uncertainty. The
lower line shows the prediction for an oscillated flux with
$\sin^2 2 \theta = 1$ and $\Delta m^2 = 0.0024$ eV$^2$.}
\end{center}
\end{figure}

The $Up$ $Through$ muon tracks we focus on in this study
come from $\nu_{\mu}$ interactions in the rock below MACRO. In these
events, the muon crosses the
entire detector requiring that $E_{\mu}> 1$ GeV.
The time information provided by the scintillator counters determines
the flight direction of the muon, allowing $Up$ $Through$ events to be distinguished from the
much more common down-going muons.
The measured muon velocity is calculated with the convention that down going muons
have $\beta$=velocity/c=+1 while up going muons have $\beta=-1$.
In the  $Up$ $Through$ event sample, almost 50\% of the tracks intercept 3
scintillators planes. In this case,  there is redundancy in the
time measurement, and $\beta$ is calculated from a linear fit of the
times as a function of the path length. Tracks with a poor fit are rejected.
 Upward going muons are selected by requiring that the measured velocity
lie in the range $-1.25\le$1/$\beta$ $\le -0.75$.

The data used in this study have been collected in three
periods, with different detector configurations, starting in 1989.
The statistics  is largely dominated by the full detector run,
started in May 1994 and ended in December 2000 (live time 5.51 years).
The total live time, normalized to the full detector configuration, is 6.17 years.

Several cuts are imposed on the data to remove backgrounds caused by
radioactivity or showering events which may result in bad
time reconstruction. The primary data selection in this regard requires
that the position of a muon hit in each scintillator, as determined from the
timing within the scintillator counter, agrees within $\pm$70 cm with
the position indicated by the streamer tube track. This eliminates events with
significant errors in timing.
In addition, downgoing muons which pass near or through
MACRO may produce low-energy, upgoing particles, which could appear to be
neutrino-induced
upward throughgoing muons if the down-going muon misses the detector
\cite{BACKSCA}.
In order to reduce this background, we impose a cut
requiring that each upgoing muon must cross at least 200 g/cm$^2$ of
material in the bottom half of the detector.
Finally, a large number
of nearly horizontal ($\cos \theta > -0.1$), but upgoing muons have
been observed coming from azimuth angles (in local coordinates) from
-30$^\circ$ to 120$^\circ$. In this direction
 the overburden is insufficient to remove
nearly horizontal, downgoing muons which have scattered in the mountain
and appear as upgoing. We exclude this region from our data.

\begin{figure}[t]
\mbox{\epsfig{file=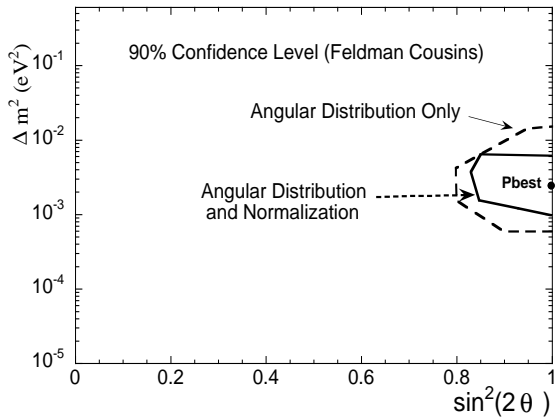,width=7.5cm,height=5.5cm}}
\caption{\label{escl}\small
The MACRO $90\%$ confidence level regions computed using the angular
distribution only (dashed line) and the angular distribution combined with
the normalization (continous line).  
}
\end{figure}

After applying the data selections described in the previous paragraph, we
observe 863 events with measured velocities in the range $-1.25 < 1/\beta <
-0.75$.
Based on events outside the upgoing muon peak, we estimate
that there are $22.5$  background events
in this data sample. In addition , we estimate
that there are $14.2$ events which result from upgoing
charged particles produced by
downgoing muons in the rock near MACRO.
Finally, it is estimated that $17$ events are the
result of interactions of neutrinos in the  bottom layer of MACRO
scintillators. After subtracting these backgrounds to the $Up Through$
data set, the number of upgoing throughgoing muons integrated over all
zenith angles is 809.

\begin{figure}[t]
\mbox{\epsfig{file=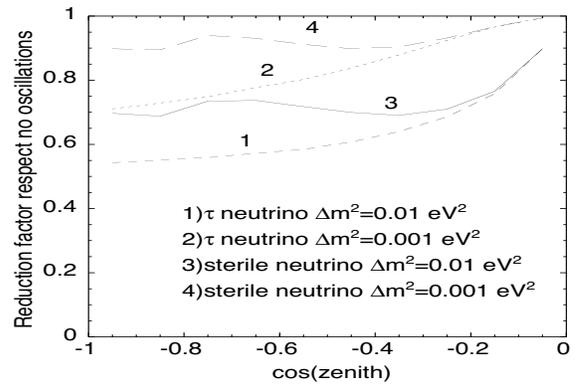,width=7.5cm,height=5.0cm}}
\caption{\label{reduction}\small
Reduction factor for  $sin^2 2\theta=1$, two values of  $\Delta m^2$  and
$\nu_{\mu}\rightarrow\nu_{s}$ or  $\nu_{\mu}\rightarrow\nu_{\tau}$ }
\end{figure}

In the simulation of our upgoing muon data,  we have used the neutrino flux
computed by the Bartol group \cite{Agrawal}, and
the GRV94 \cite{Gluck}  parton
distribution set, which increases the upgoing muon flux by +1\% with respect to the S$_1$
\cite{Morfin91} parton distribution that
 we have used in the past. For low energy channels (quasi-elastic and 1 pion
production)  we have used the cross section in
\cite{Lipari:1995pz}.
The propagation of muons to the detector has been done using
the energy loss calculation by Lohmann et al.
\cite{Lohmann85} for standard rock.
The total systematic uncertainty in the predicted flux
of upgoing muons,
 adding in quadrature the errors from the Bartol neutrino flux,
the neutrino cross-section, and muon propagation, is estimated to be $\pm17\%$.
This theoretical error in the predicted flux is mainly a scale error that does not change
the shape of the angular distribution.
Assuming no oscillations, the number of expected events integrated over
all zenith angles is
1122, giving a ratio of the observed number of events to the expectation
of 0.72 $\pm0.026$(stat) $\pm0.043$(systematic) $\pm0.12$(theoretical).

Figure~\ref{angleflux} shows the zenith angle distribution of the measured
flux of upgoing muons with energy greater than 1 GeV for our full upgoing
data sample,
compared to the Monte Carlo expectation for no oscillations,
 and with a $\nu_{\mu} \rightarrow\nu_{\tau}$ oscillated flux
with maximum mixing and
$\Delta m^2 = 0.0025$ eV$^2$.
The shape of the angular distribution has been tested with the hypothesis
of no oscillations, normalizing the total predicted flux to that observed.
The $\chi^2$ is $25.9$
for 9 degrees of freedom (P=0.2\%). Under the hypothesis of $\nu_{\mu}
\rightarrow\nu_{\tau}$ oscillation, the best $\chi^2$ is 7.1 and is
outside the physical region.
The best $\chi^2$ in the physical region of the oscillation parameters is
9.7 (P=37\%) for
$\Delta m^2$ of $0.0025$ $eV^2$ and maximum mixing.
Combining information from the angular distribution and
the  total number of events according to the procedure described in \cite{Roe},we obtain a peak probability of 66\% for
oscillations with
$\Delta m^2$ of $0.0024$ $eV^2$ and maximum mixing, while the probability for
no oscillations is 0.2\%.

The $90\%$ confidence level regions of the MACRO upgoing events are
shown in Figure~\ref{escl}. The
limits are computed using the Feldman-Cousins procedure\cite{Feldman}.
 Figure~\ref{escl} shows  the results obtained using the angular distribution
alone, and the angular distribution together with the information due to the overall normalization.
The $90\%$ confidence level regions are smaller than the regions obtained
by SuperKamiokande \cite{Fukuda:1999ah} and Kamiokande \cite{Hatakeyama:1998ea} for
the upgoing muon events.
This can be accounted for through the following effects:  
the different energy threshold (Superkamiokande has an
average energy threshold of about 7 GeV, MACRO has 1.5 GeV),
the use of the Feldman-Cousins  procedure,
and the fact that our best point is outside the
physical region.

\section{TWO FLAVORS STERILE NEUTRINO OSCILLATIONS AND TAU NEUTRINO OSCILLATIONS}
\begin{figure} 
\mbox{\epsfig{file=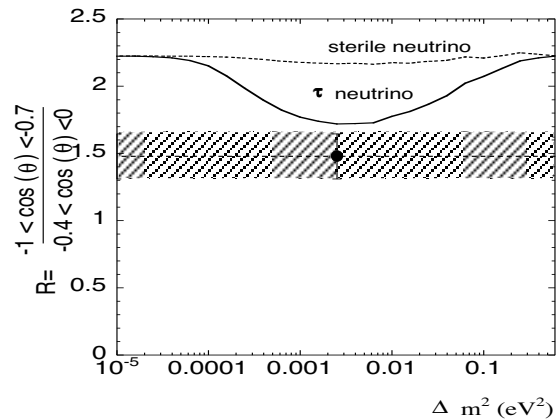,width=7.5cm,height=5.5cm}}
\caption{\label{ratio}\small The ratio between the data in two bins
(dashed line) and the comparison with the $\nu_{s}$ and $\nu_{\tau}$ oscillations
with $sin^2 2\theta=1$. The error bar includes statistical and systematical
error.}
\end{figure}

In the $\nu_{\mu}-\nu_{s}$ oscillation scenario, the matter effect changes the shape
of the angular distribution and the total number of events with respect to vacuum
oscillations.
Large matter effects are expected for neutrinos near vertical incidence,  due to the
large neutrino path length in this case, and to the increase in the density of the Earth near its core.
Assuming maximal mixing, as suggested by all available data, the matter effect
produces a reduction of the oscillation effect, and results in an upgoing muon
flux closer to that predicted by the no oscillation scenario.  This
effect would be most pronounced for directions near the vertical
 \cite{sterile}.
 Figure~\ref{reduction}
 shows the reduction with respect to no oscillations for maximal mixing
for $\nu_{\mu}\rightarrow\nu_{s}$  and $\nu_{\mu}\rightarrow\nu_{\tau}$ oscillations,
with  $\Delta m^2=0.001 eV^2$ and $\Delta m^2=0.01 eV^2$.
We have tested the shape of the observed upgoing muon angular distribution
against the hypothesis of  $\nu_{\mu}-\nu_{s}$
oscillations with maximum mixing. The
 best $\chi^2$ is $20.1$  with 9 degrees of freedom. Combining the
information obtained from the angular distribution and the normalization
the highest probability obtained is 8\% for maximum mixing
and $\Delta m^2=0.006eV^2$.
A statistically more powerful test is based on the ratio between the number of
events in the two angular regions $cos(\theta)\ge0.7$ and $cos(\theta)\le0.4$  as shown in
Figure~\ref{ratio}. This quantity is statistically more powerful than the $\chi^2$ in
10 bins because data are binned  to maximize the difference between the two
hypotheses to be tested and because the ratio is sensitive to the sign of the
variation (while the $\chi^2$  is not).
In addition, this ratio is insensitive to most of the errors in the theoretical prediction of the
$\nu$ flux and cross section.
The primary disadvantage of this statistic is the loss of  some features of the angular distribution.
We have chosen  slightly  different angular regions than suggested
in the original proposal for this statistic, presented in ref \cite{Lipari:1998}.
In our study, the angular regions used are based on
a Montecarlo study of the intervals providing the
maximum discrimination between the
$\nu_{\mu}\rightarrow\nu_{s}$  and the
$\nu_{\mu}\rightarrow\nu_{\tau}$   oscillation hypotheses.

The ratio of the flux of upgoing muons in two angular intervals is
insensitive to uncertainties in the overall $\nu$ flux and cross section,
as pointed out in the last paragraph. Several effects do, however, lead to
systematic errors in this ratio.  For example, uncertainties in
the $K$,$\pi$ fraction in atmospheric air showers, and the different angular
distributions of neutrinos produced by these parents, leads to approximately
a 3\% systematic error\cite{Lipari2000} in the predicted value for this ratio.
Another theoretical error, at the level of approximately $2\%$ for MACRO,
results from uncertainties in the neutrino cross sections, and the different
energy distributions of neutrinos arriving from the horizontal and vertical directions.
A final source of systematic error in the prediction of the flux ratio
results from the seasonal variations of the
atmosphere's density profile, and the fact that the neutrino flux is computed for
the standard United States atmosphere \cite{Agrawal} not taking into account
variations of the density profile with latitude.
Seasonal variation of the high energy muon flux has been observed by
MACRO\cite{Ambrosio:1997tc} at $42^0$ North latitude, where a $3\%$
difference was observed between summer and winter. At more extreme latitudes,
Amanda\cite{Bouchta:1999kg}, which operates near the South
Pole, observes a  $20\%$  difference between  winter and summer.
A precise estimate of the seasonal variation of the high energy neutrino flux is rather difficult
to obtain because it requires knowing the density profile of the atmosphere  over the entire Earth.
We have performed a simplified estimate of the size of this effect based on
an analytic neutrino flux calculation \cite{Lipari:1993hd} and
the CIRA-86 atmosphere tables \cite{CIRA86}. According to this calculation  the amplitude of the seasonal
variations of the ratio of the vertical to horizontal neutrino flux
is of the order of  $\pm2.6\%$. Assuming a
sinusoidal variation during the year,  this amplitude corresponds to a root mean square
value of about $1.3\%$.
 Dividing the MACRO data into a winter set (including the months from
November up to April)  and a summer set (the remaining months),
we observe a difference in the ratio of the flux in the two angular bins of $19\%\pm 17\%$ between
the two data sets, with a smaller value in the summer as expected for
the seasonal variation, compatible inside the large errors with the expectations.
 We include in our estimate of the total systematical error in the predicted
flux ratio a $1.3\%$ contribution due to  seasonal variations.
The systematic error  due to the fact that the neutrino flux is calculated
using the standard United States atmosphere has been estimated to be less than 1\%.
Accounting for all contributions to the systematic error, we estimate
that the total uncertainty in the predicted value for the flux ratio is 4\%.

The total systematic error in the measured value of the flux ratio has
been estimated to be 4.6\%. This error is  due
to uncertainties in the efficiency of the analysis cuts and detector efficiencies; it could be reduced in
the future with a  reprocessing of the data to correct for the change of the
apparatus operating conditions with  time. Combining in quadrature the theoretical error
and the experimental error we obtain a total error in the ratio of about 6\%.

In the full upgoing muon data set, there are 305 events with
$cos(\theta)\le-0.7$, and 206 events with $cos(\theta)\ge-0.4$, giving a value
for the flux ratio of $R_{exp}=1.48\pm0.13_{stat}  \pm0.10_{syst}$. This
measured value can be compared with $R{\tau}_{min}$=1.72 and
$Rsterile_{min}$=2.16, which are the minimum possible values
of $R$ for $\nu_{\mu}\rightarrow\nu_{\tau}$ and
$\nu_{\mu}-\nu_{s}$ oscillations respectively, for maximum mixing and $\Delta m^2$ of $0.0025$ $eV^2$.
For  values of $sin^2 2\theta \le1$ the value of R is larger than
$R_{min}$ both for $\nu_{\mu}-\nu_{s}$ and $\nu_{\mu}\rightarrow\nu_{\tau}$. We note
that this ratio does not have a gaussian distribution - the errors are reported only to give a crude
estimate of the statistical significance.
 The corresponding one sided
probability
$Pbest_{\tau}$ of measuring a value smaller than $R_{exp}$, assuming a true value for
the ratio of $R{\tau}_{min}$, is $8.4\%$.  For $\nu_{\mu}-\nu_{s}$
the  probability $Pbest_{ster}$ is 0.033\%.
 The ratio of the probabilities $Pbest_{\tau}$ /  $Pbest_{ster}$ is 254.  This
implies that $\nu_{\mu}-\nu_{s}$ oscillation (with any mixing) is excluded at
about 99\% C.L. compared with $\nu_{\mu}\rightarrow\nu_{\tau}$ oscillation
with maximum mixing. In calculating these confidence limits we have considered
correctly the non gaussian distribution of the ratio.

Additional information could be derived from the total number of events, at
the expense of larger theoretical uncertainties.
For the best
value of $\Delta m^2$ for sterile neutrino oscillation we expect a flux
reduction of $R_{flux}=0.83$ for  $\nu_{\mu}\rightarrow\nu_{s}$  and $\Delta
m^2=0.0025^2eV^2$, to be compared with  the measured value 0.72. However,
due to
the large theoretical uncertainty, the total number of events
was not used in the statistical analysis presented here.

It should be noted that this analysis has been carried out for
the two neutrino mixing case.
A more complicated oscillation scenario, with
3 or more neutrinos\cite{Fogli:2001ir},
or the scenario with large extra dimensions
\cite{Barbieri:2000mg} cannot be excluded.

In conclusion, using the improved statistics afforded by the full MACRO
data set, the test of the shape of the angular
distribution of upgoing muons  is in good agreement with
$\nu_{\mu}\rightarrow\nu_{\tau}$ oscillation, and maximal mixing. The
best $\chi^2$ is $9.7$  for 9 degrees of freedom.
Based on the ratio test,
the $\nu_{\mu}\rightarrow\nu_{s}$ oscillation hypothesis has a 0.033\%
probability of agreeing with the data, and is disfavored at more 
than 99\% C.L.with respect to the best fit point of
$\nu_{\mu}\rightarrow\nu_{\tau}$ oscillation.

We thank T. Kaijta for useful discussions to compare our data to those of 
Superkamiokande.
We acknowledge the staff of the {\it Laboratori Nazionali del Gran Sasso} and the invaluable
assistance of the technical staffs of all the participating Institutions.  For generous
financial contributions we thank the U.S.~Department of Energy,
the National Science Foundation, and the Italian {\it Istituto
Nazionale di Fisica Nucleare}: both for direct support and for FAI
grants awarded to non-Italian MACRO collaborators.

 \end{document}